\documentclass{amsart}
\usepackage{amssymb}
\usepackage{amsmath}

\theoremstyle{plain}
\newtheorem{theorem}{Theorem}[section]
\newtheorem{lemma}[theorem]{Lemma}
\newtheorem{proposition}[theorem]{Proposition}
\theoremstyle{definition}

\theoremstyle{remark}

\numberwithin{equation}{section}

\input{tcilatex}

\begin{document}
\title[The structure of \ Delsarte-Darboux type binary transformations]{The
\ Delsarte-Darboux type binary transformations and their
differential-geometric and operator structure. Part 1}
\author{Y.A. Prykarpatsky, A.M. Samoilenko }
\address{The Institute of Mathematics at the NAS, Kiev 01601, Ukraine, the
AGH University of Science and Technology, Department of Applied Mathematics,
Krakow 30059 Poland, }
\curraddr{ Brookhaven Nat. Lab., CDIC, Upton, NY, 11973 USA}
\email{yarpry@bnl.gov, sam@imath.kiev.ua}
\thanks{The fourth author was supported in part by a local AGH grant for
which he thanks.}
\author{A.K.~Prykarpatsky}
\address{The AGH University of Science and Technology, Department of Applied
Mathematics, Krakow 30059 Poland, and Dept. of \ Nonlinear Mathematical
Analysis at IAPMM, NAS of Ukraine, Lviv 79601 Ukraina}
\email{prykanat@cybergal.com, pryk.anat@ua.fm}
\author{V.Hr. Samoylenko}
\address{Dept. of Mchanics and Mathemmatics, National Taras Shevchenko
University, Kyiv, 00004, Ukraina }
\email{svhr@univ.kiev.ua}
\date{}

\begin{abstract}
The structure properties of multidimensional Delsarte-Darboux transmutation
operators in parametric functional spaces is studied by means of
differential-geometric and topological tools. It is shown that kernels of
the corresponding integral operator expressions depend on the topological
structure of related homological cycles in the coordinate space. As a
natural realization of the construction presented we build pairs of Lax type
commutative differential operator expressions related via a Delsarte-Darboux
transformation and having a lot of applications in soliton theory.
\end{abstract}

\maketitle

\section{Introduction: conjugated differential operators and their
properties.}

\setcounter{equation}{0}Let us consider the following functional space $%
\mathcal{H}:=L_{1}(l_{(t,y)}^{2};H),$ where $H=L_{2}(\mathbb{R}_{x};\mathbb{C%
}^{N}),$ $N\in \mathbb{Z}_{+},$ in which the next matrix-differential
expressions
\begin{eqnarray}
\frac{\partial }{\partial y}-L &:&=\frac{\partial }{\partial y}%
-\sum\limits_{i=0}^{n(L)}a_{i}(x;y,t)\frac{\partial ^{i}}{\partial x^{i}}:=%
\mathcal{L},\text{ \ }  \label{1} \\
\frac{\partial }{\partial t}-M &:&=\frac{\partial }{\partial t}%
-\sum\limits_{j=0}^{n(M)}b_{j}(x;y,t)\frac{\partial ^{j}}{\partial x^{j}}:=%
\mathcal{M},  \notag
\end{eqnarray}%
are defined. Here $l_{(t,y)}^{2}=l^{2}:=[0,Y]\times \lbrack 0,T]$ $\in
\mathbb{R}_{+}^{2},$ matrices $a_{i},b_{j}\in C^{1}(l_{(t,y)}^{2};S(\mathbb{R%
};End\mathbb{C}^{N})),$ $i=\overline{1,n(L)},$ $j=\overline{1,n(M)},$ where $%
S(\mathbb{R};End\mathbb{C}^{N})$ is the space of matrix-valued Schwartz
class coefficient functions, and $n(M),$ $n(L)\in \mathbb{Z}_{+}$ are fixed
orders. Let $\mathcal{H}^{\ast }:=L_{1}(l_{(t,y)}^{2};H^{\ast })$ be the
corresponding conjugated s to $\mathcal{H}$ \ space.

Let us define on the space $\mathcal{H}^{\ast }\times \mathcal{H}$ an
ordinary semi-linear scalar form according to the rule: for any pair $%
(\varphi ,\psi )\in \mathcal{H}^{\ast }\times \mathcal{H}$
\begin{equation}
(\varphi ,\psi ):=\int_{l_{(t,y)}^{2}}dtdy\int\limits_{\mathbb{R}%
}dx\,<\varphi ,\psi >=\int_{l_{(t,y)}^{2}}dtdy\int\limits_{\mathbb{R}}dx\,(%
\bar{\varphi}^{\intercal }\psi ),  \label{2}
\end{equation}%
where $\ <\,\cdot \,,\,\cdot \,>$ is the standard semi-linear scalar form on
$\mathbb{C}^{N},\ $the bar $\ "-"$ means the usual complex conjugation and $%
"\intercal "$ means the standard matrix transposition. Concerning the scalar
form (\ref{2}) let us study a problem of existence the corresponding to (\ref%
{1}) conjugated matrix differential operators in the space $\mathcal{H}%
^{\ast }.$

Differential expressions $L,M:H\rightarrow H$ \ are defined on the domain $%
Dom(L,M)\subset H,$ being dense in $H.$ Then, by definition, the conjugated
operators $L^{\ast },$~ $M^{\ast }:H^{\ast }\rightarrow H^{\ast }$ exist and
for all $\varphi ,\psi \in W_{1}^{1}(l_{(t,y)}^{2};Dom(L,M))$ the following
equalities
\begin{equation}
(\mathcal{L}^{\ast }\varphi ,\psi )=(\varphi ,\mathcal{L}\psi ),\quad (%
\mathcal{M}^{\ast }\varphi ,\psi )=(\varphi ,\mathcal{M}\psi )  \label{3}
\end{equation}%
obviously hold. Then one can consider the corresponding to (\ref{3})
frelationships being analogs of the classical Lagrange identities for the
operator $\mathcal{L}$
\begin{equation}
<\mathcal{L}^{\ast }\varphi ,\psi >-<\varphi ,\mathcal{L}\psi >=-\frac{%
\partial }{\partial x}\,Z_{(L)}[\varphi ,\psi ]+\frac{\partial }{\partial t}(%
\bar{\varphi}^{\intercal }\psi )  \label{4}
\end{equation}%
and for the operator $\mathcal{M}$
\begin{equation}
<\mathcal{M}^{\ast }\varphi ,\psi >-<\varphi ,\mathcal{M}\psi >=-\frac{%
\partial }{\partial x}\,Z_{(M)}[\varphi ,\psi ]-\frac{\partial }{\partial y}(%
\bar{\varphi}^{\intercal }\psi ),  \label{5}
\end{equation}%
where $Z_{(L)}[\varphi ,\psi ]$ \ and $Z_{(M)}[\varphi ,\psi ]$ are some
semi-linear forms on $\mathcal{H}^{\ast }\times \mathcal{H}$. From (\ref{4})
and (\ref{5}) one sees that the conjugated operators $L^{\ast }:H^{\ast
}\rightarrow H^{\ast }$ and \ $M^{\ast }:H^{\ast }\rightarrow H^{\ast \text{
}}$are defined if there exists a matrix $\Omega \in C^{1}(\mathbb{R}%
^{2}\times l^{2};\mathbb{C})$ satisfying the expressions
\begin{equation}
\bar{\varphi}^{\intercal }\psi :=\partial \Omega /\partial x,\quad
Z_{(L)}[\varphi ,\psi ]=\partial \Omega /\partial t,\text{ }Z_{(M)}[\varphi
,\psi ]=\partial \Omega /\partial y,  \label{6}
\end{equation}%
together with conditions
\begin{equation}
\text{ }\partial \Omega /\partial y,\text{ }\partial \Omega /\partial t\in
W_{1}^{1}(l_{(t,y)}^{2};H).  \label{7}
\end{equation}%
By means of integration (\ref{4}) and (\ref{5}) correspondingly with respect
to the measures $dt\wedge dx$ and $dy\wedge dx$ we find that a function $%
\Omega \in C^{1}(\mathbb{R}^{1}\times l^{2};\mathbb{C}),$ called a Delsarte
transmutation generator, exists if there holds the following condition: the
differential form
\begin{equation}
Z^{(1)}[\varphi ,\psi ]:=Z_{(L)}[\varphi ,\psi ]dy+Z_{(M)}[\varphi ,\psi
]dt)+\bar{\varphi}^{\intercal }\psi dx=d\Omega \lbrack \varphi ,\psi ],
\label{8}
\end{equation}%
is exact, that is one can write down the following relationship
\begin{equation}
\Omega \lbrack \varphi ,\psi ]=\Omega _{0}[\varphi ,\psi
]+\int_{S(P;P_{0})}Z^{(1)}[\varphi ,\psi ],  \label{9}
\end{equation}%
where $S(P;P_{0})$ $\subset \mathbb{R\times }l^{2}$ is some smooth curve
connecting a running point $P(x;y,t)\in \mathbb{R\times }l^{2}$ with a fixed
point $P(x_{0};y_{0},t_{0})\in \mathbb{R\times }l^{2},$ a function $\Omega
_{0}[\varphi ,\psi ]$ \ is a semilinear form on $\mathcal{H}^{\ast }\times
\mathcal{H}$ constant with respect to variables $(x;y,t)\in \mathbb{R\times }%
l^{2}.$ \ It is clear that conditions (\ref{7}) for the mapping (\ref{9})
are certain restrictions concerning $(x;t,y)$--parametric dependence of
functions $(\varphi ,\psi )\in \mathcal{H}_{0}^{\ast }\times \mathcal{H}%
_{0}. $ Let $\mathcal{H}_{0}^{\ast }\times \mathcal{H}_{0}$ be a closed
subspace of \ pairs of \ functions $(\varphi ,\psi )\in $ $\mathcal{H}%
_{-}^{\ast }$ $\times \mathcal{H}_{-}$ \ where $\mathcal{H}_{-}^{\ast }$ $%
\times \mathcal{H}_{-}$ are the correspondingly Hilbert-Schmidt rigged \cite%
{Be, BS} spaces $L_{1}(l_{(t,y)}^{2};H_{-}^{\ast })\times
L_{1}(l_{(t,y)}^{2};H_{-})$. Consider \ the expression (\ref{8}) for $%
\varphi \in \mathcal{H}_{0}^{\ast }\subset \mathcal{H}_{-}^{\ast }$ and $%
\psi \in \mathcal{H}_{0}\subset \mathcal{H}_{-}$ satisfying the conditions (%
\ref{7}). It is enoug to assume that $\mathcal{L}\psi =0$ and $\mathcal{M}%
\psi =0$ for $\psi \in \mathcal{H}_{0}$ oraz $\mathcal{L}^{\ast }\varphi =0$
and $\mathcal{M}^{\ast }\varphi =0$ for all $\varphi \in \mathcal{H}%
_{0}^{\ast },$ where
\begin{eqnarray*}
\mathcal{H}_{0} &:&=\{\psi (\lambda ;\xi )\in \mathcal{H}_{-}:\mathcal{L}%
\psi (\lambda ;\xi )=0,\text{ }\mathcal{M}^{\ast }(\lambda ;\xi )=0,\text{ }%
\psi (\lambda ;\xi )|_{\overset{t=0^{+}}{y=0^{+}}} \\
&=&\psi _{\lambda }\in H_{-}^{\ast },\text{ \ }L\psi _{\lambda }=\lambda
\psi _{\lambda },\text{ \ }\psi (\xi )|_{x=x_{0}}=0,\text{ }\ (\lambda ;\xi
)\in \Sigma \\
&=&\sigma (L,M)\cap \bar{\sigma}(L^{\ast },M^{\ast })\times \Sigma _{\sigma
}\},
\end{eqnarray*}%
\begin{eqnarray}
\mathcal{H}_{0}^{\ast } &:&=\{\varphi (\lambda ;\xi )\in \mathcal{H}%
_{-}^{\ast }:\mathcal{L}^{\ast }\varphi (\lambda ;\xi )=0,\text{ }\mathcal{M}%
^{\ast }\varphi (\lambda ;\xi )=0,\text{ }\varphi (\lambda ;\xi )|_{\overset{%
t=0^{+}}{y=0^{+}}}  \label{10} \\
&=&\varphi _{\lambda }\in H_{-}^{\ast },\text{ \ }M\varphi _{\lambda }=\bar{%
\lambda},\text{ }\varphi _{\lambda },\varphi (\xi )|_{x=x_{0}}=0,\text{ }%
(\lambda ;\xi )\in \Sigma  \notag \\
&=&\sigma (L,M)\cap \bar{\sigma}(L^{\ast },M^{\ast })\times \Sigma _{\sigma
}\}  \notag
\end{eqnarray}%
for some fixed point $x_{0}\in \mathbb{R}$ and a "specrtral" set $\Sigma
=\sigma (L,M)\cap \bar{\sigma}(L^{\ast },M^{\ast })\times \Sigma _{\sigma
}\subset \mathbb{C}^{p},$ $\sigma (L,M)\subset \mathbb{C}$ is a combined
spectrum of operators $L$ and $M$ \ and $\sigma (L^{\ast },M^{\ast })\subset
\mathbb{C}$ is that for the adjoint pair $L^{\ast }$ and $M^{\ast }$ in $H.$
Thereby one can formulate the following proposition.

\begin{proposition}
Consider a pair of functions $(\varphi ,\psi )\in \mathcal{H}_{0}^{\ast }$ $%
\times \mathcal{H}_{0};$ then there exists a semilinear form $\Omega :%
\mathcal{H}_{0}^{\ast }$ $\times \mathcal{H}_{0}\rightarrow \mathbb{C},$ \
such that relationships (\ref{8}) and (\ref{9}) hold.
\end{proposition}

The proposition above makes it possible to study now local prperties of
operators $\mathcal{L}$ and $\mathcal{M}$ in $\mathcal{H}$ withrespect to
some their dependence on parameter variables $(y,t)\in l^{2}.$ To study this
dependence let us proceed to its corresonding analysis in the next Chapter.

\section{ Structure of Delsarte-Darboux transformations}

\setcounter{equation}{0}Consider now another pair of operators $\mathcal{%
\tilde{L}}$ and $\mathcal{\tilde{M}}:\mathcal{H}\rightarrow \mathcal{H}$ for
which there exist the corresponding conjugated operators $\tilde{L}^{\ast }$
and $\tilde{M}^{\ast }:H^{\ast }\rightarrow H^{\ast }.$ Making use of the
Proposition 1 we can easily find another Delsrate transmutation generator $%
\tilde{\Omega}\in C^{1}(\mathbb{R}\times l^{2};\mathbb{C})$ being a
semilinear form on suitable pairs of functions $(\tilde{\varphi},\tilde{\psi}%
)\in \mathcal{\tilde{H}}_{0}^{\ast }\times \mathcal{\tilde{H}}_{0}$ for
which there exist the corresponding semi-linear forms $\tilde{Z}_{(\tilde{L}%
)}[{\tilde{\varphi}},{\tilde{\psi}}],$ $\tilde{Z}_{(\tilde{M})}[{\tilde{%
\varphi}},{\tilde{\psi}}]$ satisfying the conditions like (\ref{4}), (\ref{5}%
), (\ref{6}) and (\ref{7}):%
\begin{equation}
\overset{\_}{\tilde{\varphi}}^{\intercal }\tilde{\psi}:=\partial \tilde{%
\Omega}/\partial x,\quad \tilde{Z}_{(\tilde{L})}[\tilde{\varphi},\tilde{\psi}%
]=\partial \tilde{\Omega}/\partial t,\text{ }\tilde{Z}_{(\tilde{M})}[\tilde{%
\varphi},\tilde{\psi}]=\partial \tilde{\Omega}/\partial y,  \label{11}
\end{equation}%
together with conditions
\begin{equation}
\text{ }\partial \tilde{\Omega}/\partial y,\text{ }\partial \tilde{\Omega}%
/\partial t\in W_{1}^{1}(l_{(t,y)}^{2};H).  \label{12}
\end{equation}
By means analysing conditions (\ref{11}) we can similarly as before state
that a function $\tilde{\Omega}\in C^{1}(\mathbb{R}^{1}\times l^{2};\mathbb{C%
})$ exists if the differential form
\begin{equation}
\tilde{Z}^{(1)}[\tilde{\varphi},\tilde{\psi}]:=\tilde{Z}_{(\tilde{L})}[%
\tilde{\varphi},\tilde{\psi}]dy+\tilde{Z}_{(\tilde{M})}[\tilde{\varphi},%
\tilde{\psi}]dt)+\overset{\_}{\tilde{\varphi}}^{\intercal }\tilde{\psi}dx=d%
\tilde{\Omega}[\tilde{\varphi},\tilde{\psi}],  \label{13}
\end{equation}%
is exact, that is one can write down the corresponding to (\ref{13})
relationship for $\tilde{\Omega}[\tilde{\varphi},\tilde{\psi}]:=\tilde{\Omega%
}(\lambda ;\xi |\mu ;\eta )\ $\ as follows:
\begin{equation}
\tilde{\Omega}(\lambda ;\xi |\mu ;\eta )=\tilde{\Omega}_{0}(\lambda ;\xi
|\mu ;\eta )+\int_{\tilde{S}(P;\tilde{P}_{0})}\tilde{Z}^{(1)}[\tilde{\varphi}%
(\lambda ;\xi ),\tilde{\psi}(\mu ;\eta )]  \label{14}
\end{equation}%
for all pairs $(\lambda ;\xi )$ and $(\mu ;\eta )\in \Sigma ,$ where by
definition,
\begin{eqnarray}
\mathcal{\tilde{H}}_{0} &:&=\{\tilde{\psi}(\lambda ;\xi )\in \mathcal{H}%
_{-}^{\ast }:\mathcal{\tilde{L}}\tilde{\psi}(\lambda ;\xi )=0,\text{ }%
\mathcal{\tilde{M}}\tilde{\psi}(\lambda ;\xi )=0,  \label{15} \\
\text{ \ }\tilde{\psi}(\lambda ;\xi )|_{\overset{t=0^{+}}{y=0^{+}}} &=&%
\tilde{\psi}_{\lambda }\in H_{-},\text{ }\tilde{L}\text{ }\tilde{\psi}%
_{\lambda }=\lambda \psi _{\lambda },\text{ \ }\tilde{\psi}(\lambda ;\xi
)|_{x=\tilde{x}_{0}}=0,  \notag \\
\text{ }(y,t) &\in &l^{2},\text{ }(\lambda ;\xi )\in \Sigma =\sigma
(L,M)\cap \bar{\sigma}(\tilde{L}^{\ast },\tilde{M}^{\ast })\times \Sigma
_{\sigma }\}  \notag \\
\mathcal{\tilde{H}}_{0}^{\ast } &:&=\{\tilde{\varphi}(\lambda ;\eta )\in
\mathcal{H}_{-}^{\ast }:\mathcal{\tilde{L}}^{\ast }\tilde{\varphi}(\lambda
;\eta )=0,\text{ }\mathcal{\tilde{M}}^{\ast }\tilde{\varphi}(\lambda ;\eta
)=0,  \notag \\
\text{ \ }\tilde{L}^{\ast }\tilde{\varphi}_{\lambda } &=&\bar{\lambda}%
\varphi _{\lambda }\tilde{\varphi}(\lambda ;\eta )|_{\overset{t=0^{+}}{%
y=0^{+}}}=\tilde{\varphi}_{\lambda }\in H_{-}^{\ast },\text{ }\tilde{\varphi}%
(\lambda ;\eta )|_{x=\tilde{x}_{0}}=0,\text{ }  \notag \\
\text{ \ \ }(y,t) &\in &l^{2},\text{ }(\lambda ;\eta )\in \Sigma =\sigma (%
\tilde{L},\tilde{M})\cap \bar{\sigma}(\tilde{L}^{\ast },\tilde{M}^{\ast
})\times \Sigma _{\sigma }\}  \notag
\end{eqnarray}%
for some\ fixed points $\tilde{x}_{0}\in \mathbb{R}$ and the "spectral"
parameter set $\Sigma :=\sigma (\tilde{L},\tilde{M})\times \Sigma _{\sigma
}\subset \mathbb{C}^{p}.$ Assume now that there exists an isomorphic bounded
mapping $\mathbf{\Omega }:\mathcal{H}_{0}\rightleftarrows \tilde{\mathcal{H}%
_{0},}$ parametrized by pairs $(\varphi (\lambda ;\xi ),\tilde{\varphi}(\mu
;\eta )\in $ $\mathcal{H}_{0}^{\ast }\times \mathcal{\tilde{H}}_{0}^{\ast }$
and $(\psi (\lambda ;\xi ),\tilde{\psi}(\mu ;\eta ))\in $ $\ \mathcal{H}%
_{0}\times \tilde{\mathcal{H}_{0}},$ $(\lambda ;\xi )$ and $(\mu ;\eta )\in
\Sigma ,$ which we will define as
\begin{equation}
\mathbf{\Omega }:\psi (\lambda ;\xi )\rightarrow {\tilde{\psi}(\lambda ;\xi
):}=\psi (\lambda ;\xi )\cdot \Omega ^{-1}\Omega _{0},  \label{16}
\end{equation}%
where one supposes that an expression $\Omega ^{-1}:L_{2}^{\rho }(\Sigma ;%
\mathbb{C})\rightarrow L_{2}^{\rho }(\Sigma ;\mathbb{C})$ denotes the
inverse operator for the corresponding kernel $\Omega (\lambda ;\xi |\mu
,\eta )\in L_{2}^{\rho }(\Sigma ;\mathbb{C})\times L_{2}^{\rho }(\Sigma ;%
\mathbb{C}),$ where $\rho $ is some finite Borel measure on the Borel
subsets of a "spectral" parameter set $\Sigma .$ The corresponding isomorhic
bounded mapping between subspaces $\mathcal{H}_{0}^{\ast }$ and $\mathcal{%
\tilde{H}}_{0}^{\ast },$ i.e. $\hat{\Omega}^{\circledast }:\mathcal{H}%
_{0}^{\ast }\rightrightarrows \mathcal{\tilde{H}}_{0}^{\ast }$ \ such that
for any pair $(\varphi (\lambda ;\xi ),\tilde{\varphi}(\mu ;\eta ))\in
\mathcal{H}_{0}^{\ast }\times \mathcal{\tilde{H}}_{0}^{\ast }$
\begin{equation}
\mathbf{\hat{\Omega}}^{\circledast }:\varphi (\lambda ;\xi )\rightarrow {%
\tilde{\varphi}}(\lambda ;\xi )=\varphi (\lambda ;\xi )\cdot \Omega
^{\circledast ,}{}^{-1}\Omega _{0}^{\circledast },  \label{17}
\end{equation}%
where, by definition, the kernel $\Omega ^{\circledast }(\lambda ;\xi ):=%
\bar{\Omega}^{\intercal }(\lambda ;\xi )\in L_{2}^{\rho }(\Sigma ;\mathbb{C}%
)\times L_{2}^{\rho }(\Sigma ;\mathbb{C}),$ $(\lambda ;\xi )\in \Sigma .$ It
is easy to see now that the following proposition holds.

\begin{proposition}
The constructed above pair of mappings $(\mathbf{\Omega }^{\circledast },%
\mathbf{\Omega })$ is consistent, i.e. there exists such a kerenel $\tilde{%
\Omega}(\lambda ;\xi )\in L_{2}^{\rho }(\Sigma ;\mathbb{C})\times
L_{2}^{\rho }(\Sigma ;\mathbb{C}),$ that conditions (\ref{12}) and~ (\ref{13}%
) hold.
\end{proposition}

\begin{proof}
Indeed, by using expressions (\ref{8}), (\ref{13}), (\ref{16}) and (\ref{17}%
) we easily obtain that
\begin{equation*}
d\tilde{\Omega}=\Omega _{0}\Omega ^{-1}d\Omega \Omega ^{-1}\Omega
_{0}=-d\left( \Omega _{0}\Omega ^{-1}\Omega _{0}\right) ,
\end{equation*}%
whence the mapping $\tilde{\Omega}=-\Omega _{0}\Omega ^{-1}\Omega _{0}$ and
the condition $\tilde{\Omega}_{0}=-$ $\Omega _{0}$ holds.$\triangleright $
\end{proof}

Since the functional spaces $\mathcal{H}_{0}$ and $\tilde{\mathcal{H}_{0}}$
are consistent, the expressions for $\mathcal{\tilde{L}}$ $:=\mathbf{\Omega }%
\mathcal{L}\mathbf{\Omega }^{-1}$and $\mathcal{\tilde{M}}:=$ $\mathbf{\Omega
}\mathcal{M}\mathbf{\Omega }^{-1}$prove to be differential too. The \
corresponding mappings $\mathbf{\Omega }$ and $\mathbf{\Omega }^{\circledast
}:\mathcal{H\rightarrow H}$ are often called Delsarte-Darboux type
transformations and were for the first time used by Darboux \cite{MS} and in
general form was studied by Delsarte \ and J. Lions \cite{De, DL, Be, Ma, LS}%
.

Consider now the compatible pair of invertible Delstare mappings $(\mathbf{%
\Omega ,}$ $\mathbf{\Omega }^{\circledast })$ from \ the closed functional
spaces $\mathcal{H}_{0}\times \mathcal{H}_{0}^{\ast }$ $\subset \mathcal{H}%
_{-}\times \mathcal{H}_{-}^{\ast }$ to closed subspeces $\mathcal{\tilde{H}}%
_{0}\times \mathcal{\tilde{H}}_{0}^{\ast }\subset \mathcal{H}_{-}\times
\mathcal{H}_{-}^{\ast }$ \ reduced naturally upon $\ \mathcal{H}\times
\mathcal{H}^{\ast }.$ It means that the following diagram
\begin{equation*}
\begin{array}{ccccc}
& \mathcal{H} & \overset{\mathbf{\Omega }}{\rightarrow } & \mathcal{H} &  \\
\mathcal{M},\mathcal{L} & \left\downarrow {}\right. &  & \left\downarrow
{}\right. & \mathcal{\tilde{L}},\mathcal{\tilde{M}} \\
& \mathcal{H} & \overset{\mathbf{\Omega }}{\rightarrow } & \mathcal{H} &
\end{array}%
\end{equation*}%
is commutative, and consequently the relationships $\mathbf{\Omega }\cdot
\mathcal{L}=\mathcal{\tilde{L}}\cdot \mathbf{\Omega }$ \ and $\mathbf{\Omega
}\cdot \mathcal{M}=\mathcal{\tilde{M}}\cdot \mathbf{\Omega }$ hold. These
relationships connect evolution operators $L$ and $M$ \ in the whole space $%
\mathcal{H}$ with the corresponding evolution operators $\tilde{L}$ and $%
\tilde{M}.$

In order to define an exact form of mappings $\mathbf{\Omega }:\mathcal{H}%
\rightarrow \mathcal{H}$ and $\mathbf{\Omega }^{\circledast }:\mathcal{H}%
^{\ast }\rightarrow \mathcal{H}^{\ast }$ we will make use of the mappings (%
\ref{16}) and (\ref{17}) on fixed elements $(\varphi (\lambda ;\xi ),\psi
(\mu ;\eta ))\in \mathcal{H}_{0}^{\ast }\times \mathcal{H}_{0},$ $(\lambda
;\xi ),(\mu ;\eta )\in \Sigma .$. Namely, from (\ref{14}) we get that
\begin{eqnarray*}
\tilde{\psi}(\lambda ;\xi ) &=&\mathbf{\Omega }(\psi (\lambda ;\xi )):=%
\underset{\Sigma }{\int }d\rho (\mu ;\eta )\underset{\Sigma }{\int }d\rho
(\nu ;\gamma )\psi (\mu ;\eta ) \\
&&\times \Omega _{x}^{-1}(\mu ;\eta |\nu ;\gamma )\Omega _{x_{0}}(\nu
;\gamma |\lambda ;\xi ),
\end{eqnarray*}%
\begin{eqnarray}
\tilde{\varphi}(\lambda ;\xi ) &=&\mathbf{\Omega }^{\circledast }(\varphi
(\lambda ;\xi )):=\underset{\Sigma }{\int }d\rho (\mu ;\eta )\underset{%
\Sigma }{\int }d\rho (\nu ;\gamma )\varphi (\mu ;\eta )  \label{18} \\
&&\times \Omega _{x}^{\circledast ,-1}(\mu ;\eta \text{\TEXTsymbol{\vert}}%
\nu ;\gamma )\Omega _{x_{0}}^{\circledast }(\nu ;\gamma |\lambda ;\xi ).
\notag
\end{eqnarray}%
that makes it possible to define the operators $\mathbf{\Omega }:\mathcal{H}%
\rightarrow \mathcal{H}$ and $\mathbf{\Omega }^{\circledast }:\mathcal{H}%
^{\ast }\rightarrow \mathcal{H}^{\ast }$ as
\begin{eqnarray*}
\mathbf{\Omega } &:&=\mathbf{1}-\underset{\Sigma }{\int }d\rho (\mu ;\eta )%
\underset{\Sigma }{\int }d\rho (\nu ;\gamma )\tilde{\psi}(\mu ;\eta )\Omega
_{x_{0}}^{-1}(\mu ;\eta |\nu ;\gamma ) \\
&&\times \int_{P_{0}}^{P}Z^{(m-1)}[\varphi (\nu ;\gamma ),(\cdot )]
\end{eqnarray*}%
\begin{eqnarray}
\mathbf{\Omega }^{\circledast } &:&=\mathbf{1}-\underset{\Sigma }{\int }%
d\rho (\mu ;\eta )\underset{\Sigma }{\int }d\rho (\nu ;\gamma )\tilde{\varphi%
}(\mu ;\eta )\Omega _{x_{_{0}}}^{\circledast ,-1}(\mu ;\eta |\nu ;\gamma )
\label{19} \\
&&\times \int_{P_{0}}^{P}\bar{Z}^{(m-1),\intercal }[(\cdot ),\psi (\nu
;\gamma )],  \notag
\end{eqnarray}%
with $\rho $ being as before some finite Borel measure on the Borel subsets
\ of $\ $a "spectral" parameter set $\Sigma .$

Now based on expressions (\ref{19}), one can easily find the "dressed"
operators $\mathcal{\tilde{L}},\mathcal{\tilde{M}}:\mathcal{H}\rightarrow
\mathcal{H},$ and thereby \ their coefficient matrix functions subject to
the corresponding coefficients of operators $\mathcal{L},\mathcal{M}:%
\mathcal{H}\rightarrow \mathcal{H},\ $which are also called the
Darboux-Backlund transformations \cite{MS}.

Note also that the \ compatibility condition for the dressed differential
operators $\mathcal{\tilde{L}},\mathcal{\tilde{M}}$ \ is equivalent to some
system of nonlinear evolution equations in partial derivatives \ and often
this pair is called \cite{No, Ma, MS} a Lax type or a Zakharov-Shabat pair.

Consider now the structure of "dressed" operators
\begin{equation}
\mathcal{\tilde{L}}=\mathbf{\Omega }\mathcal{L}\mathbf{\hat{\Omega}}%
^{-1},\quad \mathcal{\tilde{M}}=\mathbf{\Omega }\mathcal{M}\mathbf{\Omega }%
^{-1}  \label{20}
\end{equation}%
as elements of orbits of some Volterra \ group $G_{-}$ \cite{Niz1, Niz2}. As
one can see from (\ref{21}) , these operators lie correspondingly on orbits
of elements $\mathcal{L},\mathcal{M}\in \mathcal{G}_{-}^{\ast }$ with
respect to the natural co-adjoint group action of the group of
pseudo-differential operators $G_{-}$, whose Lie co-algebra $\mathcal{G}%
_{-}^{\ast }$ consists of Volterra type integral operators of the form $%
l:=\sum\limits_{i=0}^{n(l)}a_{i}\partial ^{-1}\bar{b}_{i}^{\intercal },$
where $n(l)\in \mathbb{Z}_{+}$ is some finite number, i.e.
\begin{equation}
\mathcal{G}_{-}^{\ast }=\{l=\sum\limits_{i=0}^{n(l)}a_{i}\partial ^{-1}\bar{b%
}_{i}^{\intercal }:a_{i},b_{i}\in C^{1}(l_{(t,y)}^{2};S(\mathbb{R};End%
\mathbb{C}^{N})),\text{ }i=\overline{1,n},\text{ }n\in \mathbb{Z}_{+}\}.
\label{21}
\end{equation}%
Let us show that these orbits leave \ the space $\mathcal{G}_{-}^{\ast }$ \
invariant, i.e. the "dressed" operators $\mathcal{\tilde{L}}$ and $\mathcal{%
\tilde{M}}:\mathcal{H}\rightarrow \mathcal{H}$ under transformation (\ref{21}%
) persist to be differential with conservation of their orders. To do it let
\ us consider an arbitrary pseudo-differential operator $\mathcal{P}:%
\mathcal{H}\rightarrow \mathcal{H}$ and note that the following identity
\begin{equation}
\mathrm{Tr}\,(\mathcal{P}f\partial ^{-1}\bar{h}^{\intercal }):=(\mathcal{P}%
f,\partial ^{-1}\bar{h}^{\intercal })_{\mathcal{G}}=(h,\mathcal{P}_{+}f)_{H}
\label{22}
\end{equation}%
holds, where, by definition, $(\cdot ,\cdot )_{H}$ denotes the scalar
product in the Hilbert space $H,$
\begin{equation*}
\mathrm{Tr}(\,\cdot \,):=\int\limits_{\mathbb{R}}\,dx\underset{\partial }{\,%
\mathrm{res}}\,\mathrm{Sp}(\,\cdot \,),
\end{equation*}%
and the operation $(\ \cdot \ )_{+}$ means the projection upon the
differential part of a given pseudo-differential expression. Based on the
relationship(\ref{22}) it is easy to prove the following \cite{SP} lemma.

\begin{lemma}
A pseudo-differential operator $\mathcal{P}:\mathcal{H}\rightarrow \mathcal{H%
}$ is pure differential if and only if the following equality
\begin{equation}
(h,(\mathcal{P}\partial ^{i})_{+}f)_{H}=(h,\mathcal{P}_{+}\partial ^{i}f)_{H}
\label{23}
\end{equation}%
holds with respect to the scalar product $(\cdot ,\cdot )_{H}$ in $H$ \ for
all $i\in \mathbb{Z}_{+}$ and any dense in $\mathcal{H}^{\ast }\times
\mathcal{H}$ \ set of pairs $(f,h)\in \mathcal{H}^{\ast }\times \mathcal{H}$
\ That is the condition (\ref{23}) is equivalent to equality$\mathcal{P}_{+}=%
\mathcal{P}.$
\end{lemma}

Making use of this Lemma in the case when $\mathcal{P}:=$ $\mathcal{L}:%
\mathcal{H}\rightarrow \mathcal{H}$ and taking into consideration the
condition (\ref{23}), one gets that
\begin{equation*}
(h,(\mathcal{\tilde{L}}\partial ^{i})_{+}f)=(h,(\mathbf{\Omega }(\frac{%
\partial }{\partial t}-L)\mathbf{\Omega }^{-1}\cdot \partial ^{i})_{+}f)=
\end{equation*}%
\begin{equation*}
=(h,\frac{\partial }{\partial t}\partial ^{i}f)-(h,[(\mathbf{\Omega }_{t}%
\mathbf{\Omega }^{-1}+\mathbf{\Omega }L\mathbf{\Omega }^{-1})\partial
^{i}]_{+}f)=
\end{equation*}%
\begin{equation*}
=(h,\frac{\partial }{\partial t}\partial ^{i}f)-\mathrm{Tr}\,\{(\mathbf{%
\Omega }_{t}{\mathbf{\Omega }}^{-1}\partial ^{i}+{\mathbf{\Omega }}\ell {%
\mathbf{\Omega }}^{-1}\partial ^{i})f\partial ^{-1}\bar{h}^{\intercal }\}=
\end{equation*}%
\begin{equation*}
=(h,\frac{\partial }{\partial t}\partial ^{i}f)-\mathrm{Tr}\,\left\{ (1-{%
\tilde{\psi}}\Omega _{0}^{-1}{}\partial ^{-1}\bar{\varphi}^{\intercal
})_{t}(1+\psi \Omega _{0}^{-1}{}\partial ^{-1}\overset{\_}{\tilde{\varphi}}%
^{\intercal })\partial ^{i}+\right.
\end{equation*}%
\begin{equation}
+(1-{\tilde{\psi}}\Omega _{0}^{-1}\partial ^{-1}\bar{\varphi}^{\intercal
})L(1+\psi \Omega _{0}\partial ^{-1}\overset{\_}{\tilde{\varphi}}^{\intercal
})\partial ^{i}f\partial ^{-1}\bar{h}^{\intercal })\equiv  \label{24}
\end{equation}%
\begin{equation*}
\equiv \mathrm{Tr}\,({\tilde{L}}(\partial ^{i}f)\partial ^{-1}\bar{h}%
^{\intercal })=(h,\tilde{L}_{+}\partial ^{i}f).
\end{equation*}%
When deriving (\ref{24}) we made use of the equalities $\mathcal{L}\psi =0,$
$\mathcal{L}^{\ast }\varphi =0$ for any pair $(\varphi ,\psi )\in \mathcal{H}%
_{0}^{\ast }\times \mathcal{H}_{0}$ and the evident condition $\mathcal{L}%
_{+}=\mathcal{L}$. Thereby in accordance with lemma 1, the operator $%
\mathcal{\tilde{L}}:\tilde{\mathcal{H}}\rightarrow \tilde{\mathcal{H}}$
remains to be \ differential and, moreover, the order $ord\,\tilde{L}=ord\,L$
that follows from the definition (\ref{20}). Similarly, the same proposition
holds also for the \textquotedblright dressed\textquotedblright\ operator $%
\mathcal{\tilde{M}}:\mathcal{H}\rightarrow \mathcal{H},$ i.e. $\mathcal{%
\tilde{M}}_{+}=\mathcal{\tilde{M}}$ and $ord\,\tilde{M}=ord\,M.$ As a
conclusion from the results obtained above one can formulate the following
proposition.

\begin{proposition}
The pair of "dressed"differential operators $\mathcal{\tilde{L}},\mathcal{%
\tilde{M}}:\mathcal{H}\rightarrow \mathcal{H}$ \ of the form (\ref{20}),
obtained as a result of the Delsarte-Darboux type transformation from a
compatible Zakharov-Shabat commuting pair of differential operators $%
\mathcal{L},\mathcal{M}:\mathcal{H\rightarrow H}$ \ in the form (\ref{1})
persists to be a compatible pair of commuting differential operators in $%
\mathcal{H}$ \ preserving their differential orders. The corresponding
coefficient matrix functions of the Delsarte-Darboux transformed
differential operators $\mathcal{\tilde{L}},\mathcal{\tilde{M}}:\mathcal{H}%
\rightarrow \mathcal{H}$\ define a so-called Backlund-Darboux transformation
for the coefficient matrix functions of the initially chosen compatible pair
$\mathcal{L},\mathcal{M}:\mathcal{H\rightarrow H}$ \ of differential
operators.
\end{proposition}

From the practical point of view at proposition 3, it is clear that the
Delsarte-Darboux transformations are especially useful for construction of a
wide class of so-called soliton \cite{MS,Ni,No,AS} and algebraic solutions
to the corresponding system of nonlinear evolution differential equations,
which is equivalent to the compatibility condition for the obtained pair of
"dressed" operators (\ref{1}). A great deal of papers is devoted (see, for
example \cite{MS, AS}) to such calculations, where particular solutions of
solitons and other types were built for different evolution differential
equations of mathematical physics.

\section{General structure of Delsarte-Darboux transformations: a
dif\-fe\-ren\-ti\-al-geometric aspect}

\setcounter{equation}{0}A preliminary analysis of the Delsarte-Darboux type
transformation operators constructed above for differential operator
expressions in the case of a single variable $x\in \mathbb{R}$ shows that
its form is rather restrictive concerning a class of possible
transformations for operator differential expressions depending on two and
more variables and admitting Lax type representations\cite{Niz1,Niz2, No,
PM, Ma}. Therefore it is important to consider a nontrivial
multi-dimensional generalization of the proposed above scheme for
construction these Delsarte-Darboux type transformations. Below we will
present some short sketch of such an approach to this problem based on the
preliminary results obtained in \cite{PSS,GPSP,PSP}.

We consider as before a parametric functional space $\mathcal{H}%
:=L_{1}(l_{t};H),$ $l_{t}:=[0,T]\in \mathbb{R}_{+},$where now $H:=L_{2}(%
\mathbb{R}^{2};\mathbb{C}^{N}),$ in which there acts a (2+1)-dimensional
differential operator expression $\mathcal{L}:\mathcal{H}\rightarrow
\mathcal{H}$ of the form
\begin{eqnarray}
\mathcal{L} &=&\partial /\partial t-L(t;x,y|\partial ),  \label{25} \\
L(t;x,y|\partial ) &:&=\sum\limits_{0\leq i+j\leq n(L)}u_{ij}\frac{\partial
^{i+j}}{\partial x^{i}\partial y^{j}},  \notag
\end{eqnarray}%
with coefficients $u_{ij}\in C^{1}(l;\mathcal{S}(\mathbb{R}^{1};End\mathbb{C}%
^{N})),$ $i,j=\overline{1,n(L)}.$ Applying the same scheme as used above we
find that for the expression (\ref{25}) the standard identity
\begin{equation*}
<L^{\ast }\varphi ,\psi >-<\varphi ,L\psi >=\frac{\partial }{\partial t}(%
\bar{\varphi}^{\intercal }\psi )
\end{equation*}%
\begin{equation}
+\frac{\partial }{\partial x}Z^{(x)}[\varphi ,\psi ]+\frac{\partial }{%
\partial y}Z^{(y)}[\varphi ,\psi ]  \label{26}
\end{equation}%
holds for all pairs $(\varphi ,\psi )\in D(\mathcal{L}^{\ast })\times D(%
\mathcal{L})\subset \mathcal{H}^{\ast }\times \mathcal{H}$, where $%
Z^{(x)}[\varphi ,\psi ]$ and~ $Z^{(y)}[\varphi ,\psi ]$ are some easily
computable semi-linear forms on $\mathcal{H}^{\ast }\times \mathcal{H}.$
From (37) with respect to the oriented measure $dt\wedge dx\wedge dy$ one
gets easily that
\begin{equation*}
(<\mathcal{L}^{\ast }\varphi ,\psi >-<\varphi ,\mathcal{L}\psi >)dt\wedge
dx\wedge dy=d(\bar{\varphi}^{\intercal }\psi \wedge dx\wedge dy
\end{equation*}%
\begin{equation}
+Z^{(x)}[\varphi ,\psi ]dy\wedge dt-Z^{(y)}[\varphi ,\psi ]dx\wedge
dt):=dZ^{(2)}[\varphi ,\psi ],\text{ \ }  \label{27}
\end{equation}%
where, by definition,
\begin{equation}
Z^{(2)}[\varphi ,\psi ]=\bar{\varphi}^{\intercal }\psi \,dx\wedge
dy+Z^{(x)}[\varphi ,\psi ]\,dy\wedge dt+Z^{(y)}[\varphi ,\psi ]\,dt\wedge dx
\label{28}
\end{equation}%
is a semilinear on $\mathcal{H}_{-}^{\ast }\times \mathcal{H}_{-}$
differential 2-form on $\mathbb{R}^{2}\times l.$ Therefore for all $t\in l$
and any $(\varphi (\lambda ;\xi ),\psi (\mu ;\eta ))$ \ \ \ \ \ \ \ \ \ \ \
\ \

$\in \mathcal{H}_{0}^{\ast }\times \mathcal{H}_{0}\subset \mathcal{H}%
_{-}^{\ast }\times \mathcal{H}_{-},$ $(\lambda ;\xi ),(\mu ;\eta )\in \Sigma
,$ from a f closed subspace of the correspondingly Hilbert-Schmidt rigged
\cite{Be,BS} parametric functional spaces $\mathcal{H}_{-}^{\ast }\times
\mathcal{H}_{-}$ with $\Sigma \subset \mathbb{C}^{p}$ being some "spectral"
parameter set, \ the expression on the right-hand side of relationship (\ref%
{27}) can be made become identically zero if the conditions%
\begin{equation}
\mathcal{L}^{\ast }\varphi =0,\text{ \ \ }\mathcal{L}\psi =0\text{ \ }
\label{29}
\end{equation}%
hold on $\mathcal{H}_{0}^{\ast }\times \mathcal{H}_{0}.$ Thereby one can
define the following closed dense subspaces $\mathcal{H}_{0}^{\ast }\subset
\mathcal{H}_{-}^{\ast }\ \ $\ and $\ \mathcal{H}_{0}\subset \mathcal{H}_{-%
\text{ \ }}$similarly to (\ref{10}) as
\begin{eqnarray*}
\mathcal{H}_{0} &:&=\{\psi (\lambda ;\xi )\in \mathcal{H}_{-}:\mathcal{L}%
\psi (\lambda ;\xi )=0,\text{ }\mathcal{M}^{\ast }(\lambda ;\xi )=0,\text{ }
\\
\psi (\lambda ;\xi )|_{\overset{t=0^{+}}{}} &=&\psi _{\lambda }\in
H_{-}^{\ast },\text{ \ }L\psi _{\lambda }=\lambda \psi _{\lambda },\text{ \ }%
\psi (\xi )|_{\Gamma }=0, \\
\text{ }t &\in &l,\ (\lambda ;\xi )\in \Sigma =\sigma (L,M)\cap \bar{\sigma}%
(L^{\ast },M^{\ast })\times \Sigma _{\sigma }\},
\end{eqnarray*}%
\begin{eqnarray}
\mathcal{H}_{0}^{\ast } &:&=\{\varphi (\lambda ;\xi )\in \mathcal{H}%
_{-}^{\ast }:\mathcal{L}^{\ast }\varphi (\lambda ;\xi )=0,\text{ }\mathcal{M}%
^{\ast }\varphi (\lambda ;\xi )=0,  \label{30} \\
\text{ }\varphi (\lambda ;\xi )|_{\overset{t=0^{+}}{}} &=&\varphi _{\lambda
}\in H_{-}^{\ast },\text{ \ }M\varphi _{\lambda }=\bar{\lambda},\text{ }%
\varphi _{\lambda },\varphi (\xi )|_{\Gamma }=0,  \notag \\
\text{ }t &\in &l,\ (\lambda ;\xi )\in \Sigma =\sigma (L,M)\cap \bar{\sigma}%
(L^{\ast },M^{\ast })\times \Sigma _{\sigma }\},  \notag
\end{eqnarray}%
where we imposed, correspondingly, on the spaces $\mathcal{H}_{-}^{\ast }$
and $\mathcal{H}_{-}$ some boundary conditions at $\Gamma \subset \mathbb{R}%
^{2}$ , where $\Gamma $ is some one-dimensional smooth curve in $\mathbb{R}%
^{2}.$ Now the differential 2-form (\ref{30}) becomes closed, i.e. $%
dZ^{(2)}[\varphi ,\psi ]=0,$ that due to the Poincare lemma \cite{Go,Ca}
brings about the following equality
\begin{equation}
Z^{(2)}[\varphi (\lambda ;\xi ),\psi (\mu ;\eta )]=d\Omega ^{(1)}[\varphi
(\lambda ;\xi ),\psi (\mu ;\eta )]\text{ }  \label{31}
\end{equation}%
for some differential 1-form $\Omega ^{(1)}[\varphi (\lambda ;\xi ),\psi
(\mu ;\eta )]$ on the space $\mathbb{R}^{3}$ and all pairs $(\varphi (\xi
),\psi (\eta ))\in \mathcal{H}_{0}^{\ast }\times \mathcal{H}_{0},$ $\xi
,\eta \in \Sigma .$ Thus, the following proposition similar to one of \cite%
{Be} holds.

\begin{proposition}
Ifthe differential 2-form (\ref{31}) is closed for all pairs $(\varphi (\xi
),\psi (\eta ))\in \mathcal{H}_{0}^{\ast }\times \mathcal{H}_{0},$ $\xi
,\eta \in \Sigma ,$ and vice versa if the 2-forms $Z^{(2)}[\varphi (\lambda
;\xi ),\psi (\mu ;\eta )],$ $\ \xi ,\eta \in \Sigma ,$ are closed , then the
pair of conjugated differential operators $(L,L^{\ast })$ is adjoint with
respect to the scalar form on $H^{\ast }\times H.$
\end{proposition}

Applying now the Stokes theorem \cite{Go, Ca} for a closed 2-form (\ref{31})
on $\mathbb{R}^{2}\times l,$ we obtain that
\begin{eqnarray*}
&&\int\limits_{S^{(2)}(\sigma ^{(1)},\sigma _{0}^{(1)})}Z^{(2)}[\varphi
(\lambda ;\xi ),\psi (\mu ;\eta )] \\
&=&\int\limits_{S^{(2)}(\sigma ^{(1)},\sigma _{0}^{(1)})}d\Omega
^{(1)}[\varphi (\lambda ;\xi ),\psi (\mu ;\eta )]=\int\limits_{\partial
S^{(2)}(\sigma ^{(1)},\sigma _{0}^{(1)})}\Omega ^{(1)}[\varphi (\lambda ;\xi
),\psi (\mu ;\eta )]
\end{eqnarray*}%
\begin{eqnarray}
&=&\int\limits_{\sigma ^{(1)}}\Omega ^{(1)}[\varphi (\lambda ;\xi ),\psi
(\mu ;\eta )]-\int\limits_{\sigma _{0}^{(1)}}\Omega ^{(1)}[\varphi (\lambda
;\xi ),\psi (\mu ;\eta )]  \label{32} \\
&:&=\Omega (\lambda ;\xi |\mu ;\eta )-\Omega _{_{0}}(\lambda ;\xi |\mu ;\eta
)\text{ \ }  \notag
\end{eqnarray}%
for some piecewise imbedded smooth compact two-dimensional surface $%
S^{(2))}(\sigma ,\sigma _{0})\subset \mathbb{R}^{2}\times l$ with the
boundary $\partial S^{(2)}(\sigma ^{(1)},\sigma _{0}^{(1)})=\sigma
^{(1)}-\sigma _{0}^{(1)},$ where $\sigma ^{(1)},\sigma _{0}^{(1)}\subset
\mathbb{R}^{2}\times l$ are some closed homological one-dimensional cycles
without self-intersections parametrized correspondingly by running point $%
P(x,y;,t)\in \mathbb{R}^{2}\times l$ and \ a fixed point $%
P(x_{0},y_{0};t_{0})\in \mathbb{R}^{2}\times l.$

Making use of the surface integral (\ref{32}) and assuming that the closed
cycle $\sigma _{0}^{(1)}\subset \mathbb{R}^{2}\times l$ is fixed, one can
define the following mappings for the corresponding Delsarte-Darboux
transformations on pairs of functions $(\varphi ,\psi )\in \mathcal{H}%
_{0}^{\ast }\times \mathcal{H}_{0}:$
\begin{eqnarray*}
\tilde{\psi}(\lambda ;\xi ) &=&\mathbf{\Omega (}\psi (\lambda ;\xi )):=%
\underset{\Sigma }{\int }d\rho (\mu ;\eta )\underset{\Sigma }{\int }d\rho
(\nu ;\gamma )\psi (\mu ;\eta ) \\
&&\times \Omega ^{-1}(\mu ;\eta |\nu ;\gamma )\Omega _{0}(\nu ;\gamma
|\lambda ;\xi ),
\end{eqnarray*}%
\begin{eqnarray}
\tilde{\varphi}(\lambda ;\xi ) &=&\mathbf{\Omega }^{\circledast }(\varphi
(\lambda ;\xi )):=\underset{\Sigma }{\int }d\rho (\mu ;\eta )\underset{%
\Sigma }{\int }d\rho (\nu ;\gamma )\varphi (\mu ;\eta )  \label{33} \\
&&\times \Omega ^{\circledast ,-1}(\mu ;\eta \text{\TEXTsymbol{\vert}}\nu
;\gamma )\Omega _{0}^{\circledast }(\nu ;\gamma |\lambda ;\xi ).  \notag
\end{eqnarray}%
where the Delsarte transmutation generator expressions $\Omega (\lambda ;\xi
|\mu ;\eta )$ and $\Omega _{0}^{\circledast }(\lambda ;\xi |\mu ;\eta )\in $
$L_{2}^{\rho }(\Sigma ;\mathbf{C})\times L_{2}^{\rho }(\Sigma ;\mathbf{C}),$
$\ (\lambda ;\xi ),(\mu ;\eta )\in \Sigma ,$ are as before considered to be
nondegenarate kernels from $L_{2}^{\rho }(\Sigma ;\mathbf{C})\times
L_{2}^{\rho }(\Sigma ;\mathbf{C}).$ The following proposition concerning the
pair of spaces $\mathcal{\tilde{H}}_{0}\ni \tilde{\psi}$ and $\mathcal{%
\tilde{H}}_{0}^{\ast }\ni \tilde{\varphi}$ holds.

\begin{proposition}
The pair of functional spaces $\mathcal{\tilde{H}}_{0}^{\ast }$ and $%
\mathcal{\tilde{H}}_{0}$ consisting correspondingly of functions $(\tilde{%
\varphi},\tilde{\psi})\in \mathcal{\tilde{H}}_{-}^{\ast }\times \mathcal{%
\tilde{H}}_{-}$ defined by the expressions (\ref{33}) can be equivalently
characterized as follows:\bigskip
\begin{eqnarray}
\mathcal{\tilde{H}}_{0} &:&=\{\tilde{\psi}(\lambda ;\xi )\in \mathcal{H}%
_{-}^{\ast }:\mathcal{\tilde{L}}\tilde{\psi}(\lambda ;\xi )=0,\text{ }
\label{34} \\
\mathcal{\tilde{M}}\tilde{\psi}(\lambda ;\xi ) &=&0,\text{ \ }\tilde{\psi}%
(\lambda ;\xi )|_{t_{0}}=\tilde{\psi}_{\lambda }\in H_{-},\text{ }\tilde{L}%
\text{ }\tilde{\psi}_{\lambda }=\lambda \psi _{\lambda },\text{\ }  \notag \\
\tilde{\psi}(\lambda ;\xi )|_{\tilde{\Gamma}} &=&0,\text{ }(\lambda ;\xi
)\in \Sigma =\sigma (\tilde{L},\tilde{M})\cap \bar{\sigma}(\tilde{L}^{\ast },%
\tilde{M}^{\ast })\times \Sigma _{\sigma }\}  \notag \\
\mathcal{\tilde{H}}_{0}^{\ast } &:&=\{\tilde{\varphi}(\lambda ;\eta )\in
\mathcal{H}_{-}^{\ast }:\mathcal{\tilde{L}}^{\ast }\tilde{\varphi}(\lambda
;\eta )=0,\text{ }  \notag \\
\mathcal{\tilde{M}}^{\ast }\tilde{\varphi}(\lambda ;\eta ) &=&0,\text{ \ }%
\tilde{L}^{\ast }\tilde{\varphi}_{\lambda }=\bar{\lambda}\varphi _{\lambda },%
\text{ \ }\tilde{\varphi}(\lambda ;\eta )|_{t_{0}}=\tilde{\varphi}_{\lambda
}\in H_{-}^{\ast },  \notag \\
\text{ \ }\tilde{\varphi}(\lambda ;\eta )|_{\tilde{\Gamma}} &=&0,\text{ }%
(\lambda ;\eta )\in \Sigma =\sigma (\tilde{L},\tilde{M})\cap \bar{\sigma}(%
\tilde{L}^{\ast },\tilde{M}^{\ast })\times \Sigma _{\sigma }\}  \notag
\end{eqnarray}%
\begin{equation}
\mathcal{\tilde{H}}_{0}^{\ast }:=0\text{ \ }  \label{41b}
\end{equation}%
for some piecewise smooth curve $\tilde{\Gamma}$ $\subset \mathbb{R}^{2}.$
\end{proposition}

Based now on this Proposition, the mappings (\ref{34}) can be extended
naturally on the whole space $\mathcal{H}_{-}^{\ast }\times \mathcal{H}_{-}$
by means of the just used before classical method of variation of constants
\cite{PSP,Ca,GPSP} \ and give rise easily to the exact forms of the pair of
\ Delsarte-Darboux mapping $(\mathbf{\Omega },\mathbf{\Omega }^{\circledast
})$ upon the whole space $\mathcal{H}^{\ast }\times \mathcal{H}$ $:$
\begin{eqnarray*}
\mathbf{\Omega } &:&=\mathbf{1}-\underset{\Sigma }{\int }d\rho (\mu ;\eta )%
\underset{\Sigma }{\int }d\rho (\nu ;\gamma )\tilde{\psi}(\mu ;\eta )\Omega
_{\sigma _{0}}^{-1}(\mu ;\eta |\nu ;\gamma ) \\
&&\times \int_{S^{(2)}(\sigma ^{(1)},\sigma _{0}^{(1)})}Z^{(m-1)}[\varphi
(\nu ;\gamma ),(\cdot )]
\end{eqnarray*}%
\begin{eqnarray}
\mathbf{\Omega }^{\circledast } &:&=\mathbf{1}-\underset{\Sigma }{\int }%
d\rho (\mu ;\eta )\underset{\Sigma }{\int }d\rho (\nu ;\gamma )\tilde{\varphi%
}(\mu ;\eta )\Omega _{\sigma _{_{0}}}^{\circledast ,-1}(\mu ;\eta |\nu
;\gamma )  \label{35} \\
&&\times \int_{S^{(2)}(\sigma ^{(1)},\sigma _{0}^{(1)})}\bar{Z}%
^{(m-1),\intercal }[(\cdot ),\psi (\nu ;\gamma )],  \notag
\end{eqnarray}%
defined for some imbedded into $\mathbb{R}^{2}\times l$ piecewise smooth
two-dimensional surface $S^{(2)}(\sigma ^{(1)},\sigma _{0}^{(1)})\subset
\mathbb{R}^{2}\times l,$ spanned between two closed homological cycles $%
\sigma ^{(1)}$ and $\sigma _{0}^{(1)}\subset \mathbb{R}^{2}\times l$ as its
boundary, that is $\partial S^{(2)}(\sigma ^{(1)},\sigma _{0}^{(1)}):=\sigma
^{(1)}-\sigma _{0}^{(1)}.$ It is seen from (\ref{35}) that found above
Delsarte transmutation operators $\mathbf{\Omega :}\mathcal{H\rightarrow H}$
and $\mathbf{\Omega }^{\mathbf{\circledast }}:\mathcal{H}^{\ast }\mathcal{%
\rightarrow H}^{\ast \text{ \ }}$ are bounded of Volterra type integral
operators, strongly depending on a measure \ $\rho $ on the "spectral"
parameter space $\Sigma $ and some piecewise smooth two-dimensional surface $%
S^{(2)}(\sigma ^{(1)},\sigma _{0}^{(1)})$ parametrized by a running point $%
P(x,y;t)\in \mathbb{R}^{2}$ $\times l$ and a fixed point $%
P(x_{0},y_{0};t_{0})\in \mathbb{R}^{2}\times l.$

Making now use of the bounded Delsarte-Darboux integral transformation
operators (\ref{35}) of Volterra type, one can now as before to construct
the \ corresponding Delsarte-Darboux transformed differential operator $%
\mathcal{\tilde{L}}:\mathcal{H}\rightarrow \mathcal{H}$ \ as follows:
\begin{equation}
\mathcal{\tilde{L}}=\mathcal{L}+[\mathbf{\Omega },\mathcal{L}]\mathbf{\Omega
}^{-1}.  \label{36}
\end{equation}%
Since the expression (\ref{36}) contains the inverse integral operator $%
\mathbf{\Omega }^{-1}:\mathcal{H}\rightarrow \mathcal{H},$ it can be found
from (\ref{35}) making use of the symmetry properties between closed
subspaces $\mathcal{H}_{0}^{\ast }\times \mathcal{H}_{0}$ and $\mathcal{%
\tilde{H}}_{0}^{\ast }\times \mathcal{\tilde{H}}_{0}$ :
\begin{eqnarray*}
\mathbf{\Omega }^{-1} &:&=\mathbf{1}-\underset{\Sigma }{\int }d\rho (\mu
;\eta )\underset{\Sigma }{\int }d\rho (\nu ;\gamma )\psi (\mu ;\eta )\tilde{%
\Omega}_{_{0}}^{-1}(\mu ;\eta |\nu ;\gamma ) \\
&&\times \int_{S^{(2)}(\sigma ^{(1)},\sigma _{0}^{(1)})}\tilde{Z}[\tilde{%
\varphi}(\nu ;\gamma ),(\cdot )]
\end{eqnarray*}%
\begin{eqnarray}
\mathbf{\Omega }^{\circledast ,-1} &:&=\mathbf{1}-\underset{\Sigma }{\int }%
d\rho (\mu ;\eta )\underset{\Sigma }{\int }d\rho (\nu ;\gamma )\varphi (\mu
;\eta )\tilde{\Omega}_{_{_{0}}}^{\circledast ,-1}(\mu ;\eta |\nu ;\gamma )
\label{37} \\
&&\times \int_{S^{(2)}(\sigma ^{(1)},\sigma _{0}^{(1)})}\overset{\_}{\tilde{Z%
}}^{(m-1),\intercal }[(\cdot ),\tilde{\psi}(\nu ;\gamma )],  \notag
\end{eqnarray}%
for $(\tilde{\varphi}(\lambda ;\xi ),\tilde{\psi}(\mu ;\eta ))\in \mathcal{%
\tilde{H}}_{0}^{\ast }\times \mathcal{\tilde{H}}_{0},$ $(\lambda ;\xi ),(\mu
;\eta )\in \Sigma ,$ satisfying the conditions (\ref{34}) .

As a results of direct calculations in (\ref{36}) based on expressions (\ref%
{37}) one can find the corresponding Delsarte-Darboux transformed
coefficient functions of the transformed operator $\mathcal{\tilde{L}}:%
\mathcal{H}\rightarrow \mathcal{H}$ parametrized by piecewise smooth closed
one-dimensional homological cycles $\sigma ^{(1)},\sigma _{0}^{(1)}\subset
\mathbb{R}^{2}\times l.$ We don't present here these expressions in the
general case of operator (\ref{25}) ) as they are too cumbersome for writing
down. Application of the constructions developed in the article we are going
to deliver in detail in Part 2.

\section{Acknowledgements}

Authors are cordially thankful to prof. Nizhnik L.P. (Kyiv, Inst. of Math.at
NAS), prof. T. Winiarska (Krakow, PK), profs. A. Pelczar and J. Ombach
(Krakow, UJ), prof. St. Bryzchczy (Krakow, AGH) and prof. Z. Peradzynski
(Warszawa, UW) for valuable discussions during their seminars of many
aspects related with problems studied in the work.

\bigskip

\bigskip

\bigskip

\bigskip

\end{document}